\documentclass[aps,prb,reprint,superscriptaddress,showpacs,showkeys,floatfix]{revtex4-1}

\usepackage{color}
\usepackage{graphicx}% Include figure files
\usepackage{dcolumn}% Align table columns on decimal point
\usepackage{bm}% bold math
\usepackage{textcomp}
\usepackage{verbatim}
\usepackage{upgreek}
\usepackage{comment}
\usepackage{hyperref}% add hypertext capabilities
\newcommand{\n}{17}

\newcommand{\ttt}{498}

\newcommand{\pdt}{127}
\newcommand{\SDpdt}{12}
\newcommand{\epdt}{3}

\newcommand{\tc}{365}

\newcommand{\pdc}{113}
\newcommand{\SDpdc}{8}
\newcommand{\epdc}{2}

\newcommand{\pdd}{101}
\newcommand{\SDpdd}{8}
\newcommand{\epdd}{2}

\newcommand{\TIEpdt}{125}

\newcommand{\TIEepdt}{4}

\newcommand{\TIEpdc}{112}

\newcommand{\TIEepdc}{3}

\begin{document}

\title{The Magnetic Structure of Individual Flux Vortices in Superconducting MgB$_2$ Derived using Transmission Electron Microscopy}

\author{J.C. Loudon}
\email{j.c.loudon@gmail.com}

\author{C.J. Bowell}
\altaffiliation[Now at ]{Integrated Plasmonics, 2122 Bryant Street, San Francisco, CA 94110, USA.}%Lines break automatically or can be forced with \\

\affiliation{Department of Materials Science and
  Metallurgy, University of Cambridge, Pembroke Street, Cambridge CB2
  3QZ, United Kingdom}

\author{N.D. Zhigadlo}

\author{J. Karpinski}
\affiliation{Laboratory for Solid State Physics, ETH Zurich, Schafmattstrasse 16, CH-8093, Zurich, Switzerland}

\author{P.A. Midgley}
\affiliation{Department of Materials Science and Metallurgy, University of Cambridge, Pembroke Street, Cambridge CB2 3QZ, United Kingdom}

\date{\today}

\begin{abstract}
Images of flux vortices in superconductors acquired by transmission
electron microscopy should allow a quantitative determination of their
magnetic structure but so far, only visual comparisons have been made
between experimental images and simulations. Here, we make a
quantitative comparison between Fresnel images and simulations based
on the modified London equation to investigate the magnetic structure
of flux vortices in MgB$_2$. This technique gives an absolute,
low-field ($\sim 30$~Oe) measurement of the penetration depth from
images of single vortices. We found that these simulations gave a good
fit to the experimental images and that if all the other parameters in
the fit were known, the penetration depth for individual vortices
could be measured with an accuracy of $\pm 5$~nm. Averaging over \n{}
vortices gave a penetration depth of
$\Lambda_{ab}=\pdc{}\pm\epdc{}$~nm at 10.8~K assuming that the entire
thickness of the sample was superconducting. The main uncertainty in
this measurement was the proportion of the specimen which was
superconducting. Allowing for a non-superconducting layer of up to
50~nm thickness on the specimen surfaces gave a penetration depth in
the range $\Lambda_{ab}=100$--115~nm, close to values of $90\pm 2$~nm
obtained by small-angle neutron scattering and 118--138~nm obtained by
radio-frequency measurements. We also discuss the use of the transport
of intensity equation which should, in principle, give a
model-independent measure of the magnetic structure of flux vortices.
\end{abstract}
%Need to say why getting a measure from individual vortices is good and comment on getting absolute numbers.

\pacs{74.25.Uv, 68.37.Lp}
% PACS, the Physics and Astronomy Classification Scheme.
\keywords{MgB$_2$, Flux vortices, Lorentz microscopy, Superconductivity.}
%Use showkeys class option if keyword display desired

\maketitle

\section{Introduction}

Superconductors expel magnetic flux from their interiors (the Meissner
effect) but if a magnetic field is applied to a type-II superconductor
which is larger than the lower critical field, $H_{c1}$, magnetic flux
penetrates the superconductor by flowing along channels called flux
vortices. Each vortex carries a single quantum of magnetic flux given
by $\Phi_0=h/2e$ where $h$ is Planck's constant and $e$ the electron
charge. They consist of a core where superconductivity is suppressed
with a radius given by the coherence length, $\xi$, surrounded by
circulating supercurrents which persist over a radius given by the
penetration depth, $\Lambda$. When flux vortices move, energy is lost
and so the performance of almost all superconducting devices is
determined by the behaviour of flux vortices.

\begin{figure}
\includegraphics[height=55mm]{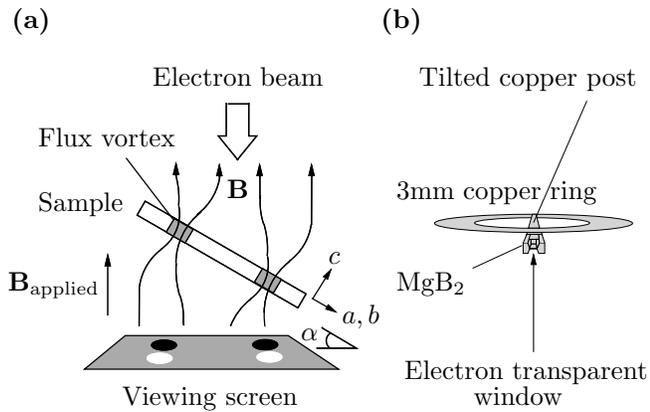}
\caption{\label{fig1} (a) Experimental arrangement for imaging flux
  vortices. The electrons are deflected by the component of the
  B-field from the vortices normal to the electron beam giving a
  black-white feature in an out-of-focus image. (b) The specimen
  geometry. The MgB$_2$ specimen was mounted to a copper post glued to a
  standard 3~mm diameter copper ring at an angle of $45^{\circ}$. An
  electron transparent window was then cut by focussed ion-beam
  milling.}
\end{figure}

Flux vortices can be imaged using transmission electron
microscopy~\cite{Harada92,Loudon12} as shown in
Fig.~\ref{fig1}(a). The superconductor is thinned to about 250~nm so
that it is electron-transparent and the flux vortices penetrate normal
to the thin surface. It is mounted at a high angle, $\alpha$
(typically 45$^\circ$), to provide a component of the B-field normal
to the electron beam so that the electrons are deflected by the
Lorentz force and appear as black-white features in an out-of-focus
image. Electron microscopy is the only imaging technique which gives
information on the internal magnetic structure of flux vortices, not
just the stray fields, and it allows imaging at video rate. As we show
here, it can be used to investigate the magnetic structure of
individual flux vortices and give an absolute measure of the
penetration depth at low magnetic fields (typically 30~Oe)
irrespective of whether the vortices form a regular array or
not. Obtaining an absolute value for the penetration depth is
important as it gives information on the number density of electrons
involved in superconductivity~\cite{Tinkham96}, the nature of the
superconducting state~\cite{Moshchalkov09} and the types of vortex
interaction which can occur~\cite{Chaves11}.

The ability to measure the magnetic properties of individual vortices
is useful for the investigation of materials in which the vortices do
not form a regular array: an effect which is hampering current
research into some of the iron-based
superconductors~\cite{Eskildsen11}. The technique would also provide a
good method to test predictions that vortices containing non-integer
multiples of the flux quantum should occur in 2-component
superconductors such as MgB$_2$ when their size becomes comparable to
the coherence length~\cite{Chibotaru10,Geurts10}. It is also useful if
the structure of certain vortices is altered by pinning as it would
then be possible to see how the vortices respond to different types of
pinning site. This type of experiment has been undertaken by Beleggia
{\it et al.}~\cite{Beleggia02} but so far the only comparisons with
theory have been visual. As shown in ref.~\onlinecite{Loudon09}, much
of the information on the structure of the vortex is contained in the
contrast of the image rather than its visual appearance. Here we use
images of flux vortices taken from MgB$_2$ to make a quantitative
comparison with simulations to assess how much information can be
obtained using transmission electron microscopy. We first compare
simulations of defocussed images with those obtained experimentally
and then discuss the use the transport of intensity equation which in
principle allows the magnetic structure of flux vortices to be derived
directly from the experimental images.

%%%%%%%%%%%%%%%%%%%%%%%%%%%%%%%%%%%%%%%%

\section{Magnesium Diboride}

MgB$_2$ was discovered to be a superconductor in
2001~\cite{Nagamatsu01}. It has a transition temperature $T_c=39$~K
and a hexagonal crystal structure (space group 191: $P6/mmm$) composed
of alternating layers of magnesium and boron with lattice parameters
$a=b=3.086$~\AA~and $c=3.542$~\AA. It is an unusual superconductor as
it has two sources of electrons which contribute to superconductivity:
one associated with the $\sigma$ bonding from the boron $p_{xy}$
orbitals and the other associated with $\pi$ bonding from the boron
$p_z$ orbitals. As a result, the penetration depth varies with the
applied field as this changes the proportion of $\sigma$ to $\pi$
electrons contributing to
superconductivity~\cite{Karmakar10,Cubitt03}. In this experiment, the
specimen was thinned to electron transparency in the $c$ direction and
since flux vortices penetrate parallel to the thin direction of the
crystal, we make a low-field ($\sim 30$~Oe) measurement of the
penetration depth in the $ab$-plane, $\Lambda_{ab}$.

Only a few papers give values for the low-field penetration depth for
MgB$_2$ resolved in crystallographic directions. Cubitt {\it et
  al.}~\cite{Cubitt03} use small-angle neutron scattering in which
measurements were extrapolated to zero field (the lowest was made at
0.1~T) to give $\Lambda_{ab}=82\pm 2$~nm at 2~K. Very low field ($\sim
1$~$\upmu$T) radio-frequency measurements were also made by Manzano
{\it et al.}~\cite{Manzano02} giving $\Lambda_{ab}=110$--130~nm and
$\Lambda_{c}=210$--280~nm and a coherence length $\xi_{ab}=5.5$~nm as
the temperature approaches absolute zero. They also show that that
$\Lambda_{ab}$ increases by $8.0\pm 1.6$~nm as the temperature is
increased from 1.35~K to 10.8~K, the temperature at which the
experiments were conducted here.

%%%%%%%%%%%%%%%%%%%%%%%%%%%%%%%%%%%%%%%%%%%%%%%

\section{Methods for Imaging Flux Vortices using Electron Microscopy}
\label{Fres_or_holo}

As flux vortices are magnetic objects, they deflect but do not absorb
the electron beam from the transmission electron microscope and so
affect only the phase and not the intensity of the electron
wavefunction. Consequently they are invisible in an in-focus
image. There are two main imaging modes by which flux vortices can be
visualised, giving access to the phase information. The first is
off-axis holography~\cite{Bonevich99} where a positively charged wire
(called an electron biprism) is placed beneath the specimen and used
to interfere electrons which passed through the specimen with those
which passed through vacuum to produce an interference pattern called
a hologram. From this, the phase of the electron wavefunction can be
calculated directly and differentiating the phase gives the B-field or
more precisely, the induction-thickness product: the component of the
B-field normal to the electron beam integrated along the path of the
beam.

The other method is out-of-focus (also called Fresnel) imaging where
images are taken out of focus and the vortices appear as black-white
features~\cite{Harada92}. These can be compared with simulations or,
if several images are taken at different defoci, the phase shift can
be reconstructed using the transport of intensity equation but this is
not as direct as holography as the intensity in a out-of-focus image
is not sensitive to the phase or the phase gradient but only to the
curvature and higher derivatives of the phase. As the
induction-thickness product is proportional to the first derivative of
the phase, this has the consequence that it can only be determined to
within an additive constant. It might be thought, therefore, that
off-axis holography would be the better method for investigating the
structure of flux vortices as it gives a direct measurement of the
phase shift but in fact, it has only been used to examine vortices in
niobium by Bonevich {\it et al.}~\cite{Bonevich93,Bonevich99}.

The main reason why out-of-focus imaging is the preferred method for
investigating flux vortices lies in the way Shot noise in the image is
transferred to noise in the recovered phase. In the Supplemental
Information~\cite{supp} we show that if the $m$-th derivative of the
phase is denoted $\phi^{(m)}$ then the noise associated with the phase
derivatives recovered by holography is

\begin{equation}
\sigma_{\phi^{(m)}_{\text{holo}}}={1 \over X^m}\sqrt{2^{m+1} \over NV^2}
\end{equation}

where $N$ is the number of counts per reconstructed pixel, $V$ is the
visibility of the holographic fringes (the difference between the
maximum and minimum intensity divided by the sum) and $X$ is the size
of one pixel.

On the other hand, the noise associated with the phase derivatives
obtained from Fresnel images is

\begin{equation}
\sigma_{\phi^{(m)}_{\text{Fresnel}}}={\pi \over X^{m-2}\lambda\Delta f}\sqrt{2^{m-1} \over N}
\end{equation}

Using reasonable values of $V=30\%$, $N=200$, $\lambda=0.00197$~nm
(the wavelength of 300~kV electrons) gives the noise in the curvature
of the phase recovered by off-axis holography as
$\sigma_{\phi''_{\text{holo}}}=2.3$~mrad/nm$^2$ for the pixel size
$X=16.9$~nm used in this experiment. The noise can be reduced by
increasing the pixel size and if $X=50$~nm is used for the same number
of electrons arriving at the detector, the noise is
$\sigma_{\phi''_{\text{holo}}}=0.09$~mrad/nm$^2$. This is close to the
lowest resolution at which individual vortices can be resolved as the
penetration depth of niobium is 52~nm. 

Simulations for a 250~nm thick specimen tilted to $\alpha=50^\circ$
give a maximum phase-curvature of 0.47~mrad/nm$^2$ for niobium (taking
$\Lambda=52$~nm~[\onlinecite{Annett04}]) and 0.12~mrad/nm$^2$ for
MgB$_2$ (taking $\Lambda=110$~nm) so it is clear that Bonevich {\it et
  al.} were working near to the limit of what is possible using
electron holography, as they acknowledge~\cite{Bonevich93}. We acquired
electron holograms from MgB$_2$ and found that although the average
B-field from the array of flux vortices could be observed, individual
vortices could not be identified.

On the other hand, the noise expected in a phase reconstruction from
defocussed images is
$\sigma_{\phi''_{\text{Fresnel}}}=0.03$~mrad/nm$^2$ for $\Delta
f=1.1$~cm and $X=16.9$~nm (used in this experiment) which is an
improvement of 2 orders of magnitude compared with holography at the
same resolution. This shows why out-of-focus imaging is the more
commonly used method for imaging flux vortices. This mode of imaging
is often referred to as `semi-quantitative' but, as we show here,
provided that the images are energy-filtered and recorded on a medium
with a linear response, they contain all the information needed to
make a quantitative comparison with simulations.

%%%%%%%%%%%%%%%%%%%%%%%%%%%%%%%%%%%%%%%%%%

\section{Experimental Methods}
\label{methods}

MgB$_2$ single crystals were synthesised by Dr J. Karpinski via the
peritectic decomposition of MgNB$_9$ as described in
ref.~\onlinecite{Karpinski03}. The samples were thinned to 250~nm in
the $c$-direction so that they were electron transparent using a
Helios Nanolab dual-beam focussed ion beam microscope. To tilt the
sample to a high angle ($\alpha$ in Fig.~\ref{fig1}(a)), the specimen
was mounted on a tilted copper post attached to a standard 3~mm copper
ring as illustrated in Fig.~\ref{fig1}(b).

Microscopy was undertaken using a Philips CM300 transmission electron
microscope operated at 300~kV equipped with an electron biprism for
holography, a `Lorentz' lens and a Gatan imaging filter. The
microscope was operated in low-magnification mode at a nominal
magnification of 105$\times$ with the main objective lens turned off
and the image was focussed with the diffraction lens. This proved more
convenient than imaging with the Lorentz lens as a wider range of
defoci could be accessed and in this mode, the selected area apertures
become objective apertures and {\it vice versa} whereas in Lorentz
mode, neither set of apertures is in the correct plane. The images
were energy filtered so that only electrons which had lost 0--20~eV on
passing through the specimen contributed to the image and an aperture
was used so that only the 000 beam and the low-angle scattering from
the vortices contributed to the image and the other crystallographic
beams were excluded. The sample was cooled using a Gatan liquid-helium
cooled `IKHCHDT3010-Special' tilt-rotate holder which has a base
temperature of 10~K.

The defocus and magnification were calibrated by acquiring images with
the same lens settings as the original images from Agar Scientific's
`S106' calibration specimen which consists of lines spaced by 463~nm
ruled on an amorphous film. The defocus was found by taking digital
Fourier transforms of the images acquired from the calibration
specimen and measuring the radii of the dark rings which result from
the contrast transfer function~\cite{Williams96}.

A thickness map of the specimen was created by dividing an unfiltered
image by an energy-filtered image and taking the natural logarithm
\cite{Egerton09} which gives the thickness parallel to the electron
beam, $l$, as a multiple of the inelastic mean free path,
$\lambda_i$. To determine $\lambda_i$, an electron hologram was taken
at room temperature at an edge of the specimen which gives a phase
shift proportional to the thickness, $\phi=C_EV_0l$. $C_E$ is a
constant which depends only on the microscope voltage and has the
value $6.523 \times 10^6$~m$^{-1}$V$^{-1}$ at 300~kV. $V_0$, the mean
inner potential, was calculated as $V_0=17.71$~V from theoretical
scattering factors given in ref.~\onlinecite{Rez94}, giving
$\lambda_i=152 \pm 2$~nm and the thickness, $l$, varied from
330--400~nm across the field of view of
Fig.~\ref{defcomp7}(a). (Ideally the thickness of the whole specimen
would have been determined by electron holography but the field of
view of the interference region was not sufficiently large.)

%%%%%%%%%%%%%%%%%%%%%%%%%%%%%%%%%%%%%%%%%%%%%%%%%%%

\section{Comparison of Defocussed Images with Simulations}
\label{comdefsim}

We first compare defocussed images with simulations based on the
London model~\cite{Tinkham96} of flux vortices. The London model
describes the B-field associated with a flux vortex but gives an
unphysical divergence of the B-field as the centre of the vortex is
approached. This is corrected by the Clem model~\cite{Clem75} where
the B-field profile is rounded near the centre of the vortex in a
region called the vortex core which has a radius given by the
coherence length, $\xi$. The London model should be a good
approximation for MgB$_2$, however, as $\xi_{ab}$ is about 6~nm
whereas the penetration depth is about 100~nm~[\onlinecite{Manzano02}] and the
pixel size of the images acquired here was 16.9~nm.

\begin{figure}
\includegraphics[width=85mm]{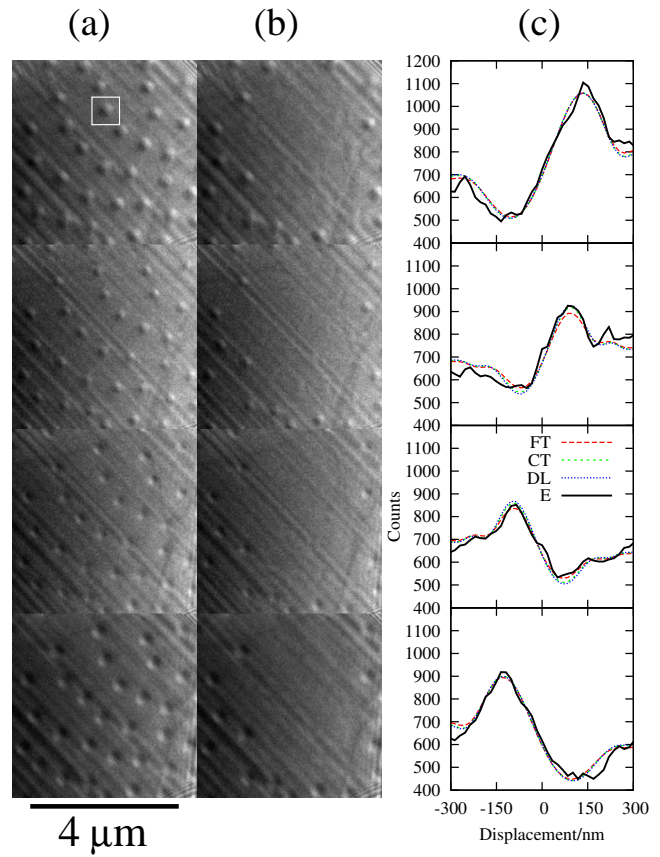}
\caption{\label{defcomp7}(a) Defocus series taken in a field of 27~G
  at 10.8~K. The defoci are (top to bottom) $\Delta f=2.24$, 1.10,
  -1.10 and -2.07~cm. (b) The same defocus series with simulated
  vortices subtracted for the case where all the variable parameters
  are optimised. The other fits mentioned in the text give very
  similar images. (c) Linescans at each defocus taken perpendicular to
  the axis of the vortex indicted by the white box in (a). FT: fitted
  thickness -- simulation in which the thickness was used as a
  variable parameter giving $l=543$~nm, $\alpha=58^\circ$,
  $\Lambda_{ab}=126$~nm. CT: calibrated thickness -- simulation in
  which the thickness was fixed at its calibrated value, $l=372$~nm,
  giving $\alpha=51^\circ$, $\Lambda_{ab}=107$~nm. DL: dead layer --
  simulation for which the thickness was fixed at $l=300$~nm to mimic
  a dead-layer of non-superconducting material, giving
  $\alpha=49^\circ$, $\Lambda_{ab}=97$~nm. E -- experimental data.}
\end{figure}

The phase shift, $\phi(x,y)$, an electron experiences on passing
through a London vortex has been calculated by Beleggia and
Pozzi~\cite{Beleggia01} using a coordinate system in which the
electron beam approaches the specimen in the $z$ direction and $x,y$
denote the image coordinates. It should be noted that their equation
correctly accounts for the boundary conditions on the B-field at the
interface between the superconductor and the vacuum so that the
spreading of the B-field near the interface and the effect of the
stray field are both contained in the model. If the intensity in the
in-focus image is uniform over the field of view (a good approximation
here since the in-focus intensity varies by about 5\% over the size of
each vortex image whereas the out-of-focus intensity varies by about
40\% at the lowest defocus used here as can be seen from
Fig.~\ref{defcomp7}(c)), the in-focus electron wavefunction can be
written as $\psi=e^{i\phi}$ and from this, any image which could be
taken using the electron microscope can be simulated.

A defocussed image can be simulated by using the Fresnel-Kirchoff
integral~\cite{Beleggia01} to propagate the in-focus electron
wavefunction by a distance $\Delta f$, called the defocus. This is
most easily accomplished by Fourier transforming the in-focus
wavefunction in $x$ and $y$ but not $z$ and multiplying by the phase
factor $e^{-i\pi\Delta f\lambda k^2}$ (where $\lambda$ is the electron
wavelength and $k$ the spatial frequency), inverse transforming and
taking the square modulus. (Here and throughout we use the Fourier
transform convention $\widetilde{f}(k)=\int_{-\infty}^\infty
f(x)e^{-2\pi ikx}\text{d}x$.) It should be noted that under the
conditions which vortices are imaged, the contribution from the
spherical aberration of the lens is negligible. Here the diffraction
lens was used to focus the image which has a spherical aberration
coefficient of several metres but the coefficient would need to be
$\sim 10^6$~m before the aberration and defocussing terms were
comparable.

The variable parameters used to model the images were: the penetration
depth $\Lambda_{ab}$, the position of the vortex in the image $x$,
$y$, the thickness of superconducting material $t$, the tilting angle
of the specimen $\alpha$ and the angle $\theta$ of the axis of the
vortex as seen in the image. These were chosen to minimise the
sum-square difference between the experimental and simulated images
using the simplex algorithm given in ref.~\onlinecite{Press92}. A
similar method was used to investigate the properties of pn junctions
in semiconductors by Twitchett {\it et al.}~\cite{Twitchett06}.

The position of each vortex $x,y$ and the rotation angle of the
vortices $\theta$ given by the simplex algorithm are readily checked
from the images. To give an independent measure of $\alpha$, the angle
between the plane of the copper ring and the specimen was measured in
the focussed-ion beam microscope as $55\pm 1^\circ$. However, in the
electron microscope, the specimen was in shadow when the ring was
horizontal and a tilt of $-8.85^\circ$ was required to reveal the
electron transparent window. This would give $\alpha=46^\circ$ if the
axes of rotation were the same but since it is likely they were
different by several degrees, $\alpha$ will be larger and we estimate
$\alpha=50\pm 5^\circ$.

The specimen thickness parallel to the electron beam, $l$, was
measured to $\pm 2$~nm using a combination of electron holography and
energy-filtered imaging as explained in section~\ref{methods}. The
thickness of the specimen parallel to the thin direction, $t$, is
related to the thickness parallel to the electron beam, $l$, via
$t=l\cos\alpha$ and the thicknesses quoted here are those parallel to
the electron beam unless otherwise stated. This method gives the total
thickness of the specimen but does not account for the possibility of
`dead-layers' of non-superconducting material produced by ion-beam
thinning at the surface of the specimen. This is the largest of the
uncertainties in this experiment and its effect is discussed later.

%alpha: Lab book 6/11/12 p189

We began by fitting all the parameters to the experimental data.
Fig.~\ref{defcomp7}(a) shows the defocus series and the black-white
circular objects are the flux vortices. The diagonal stripes in the
image are thickness undulations unintentionally introduced by focussed
ion beam milling. As discussed in ref.~\onlinecite{Loudon12} which
used the same sample, they represent thickness undulations of about
1~nm. Vortices which were strongly affected by the contrast generated
by these undulations were excluded from the fit. (b) shows the same
images but where the fitted vortices have been subtracted. The fact
that the vortices have been very effectively removed indicates that
the fit is good. (c) shows linescans at each defocus taken
perpendicular to the axis of a representative vortex indicated in (a)
by the white box and it can be seen that there is a good match between
simulation and experiment.

\begin{figure}
\includegraphics[width=87mm]{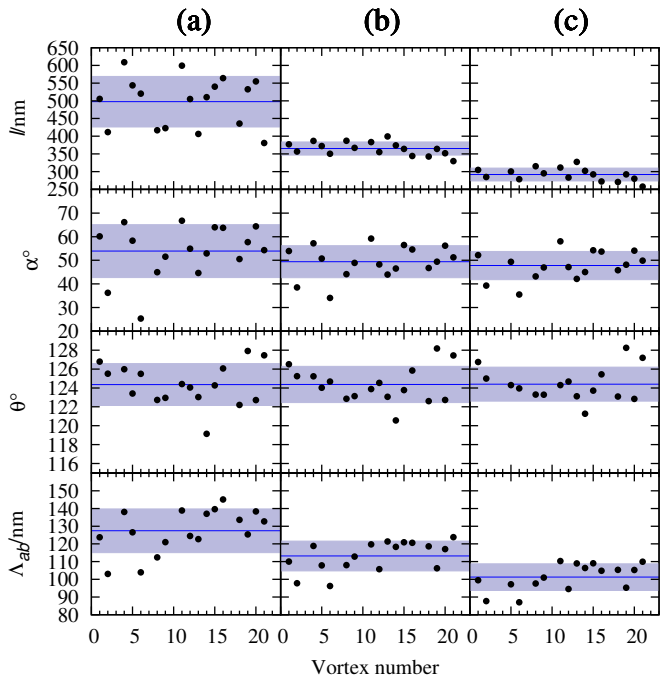}
\caption{\label{defcomparison7} Scatter graphs showing the results of
  the parameter optimisation generated by the simplex algorithm. In
  (a), the specimen thickness parallel to the electron beam, $l$, was
  treated as a variable parameter. In (b), it was fixed at its
  calibrated value. In (c), the thickness was reduced by 50~nm (72~nm
  parallel to the electron beam) from its calibrated value to simulate
  a `dead-layer' of non-superconducting material. The horizontal line
  shows the mean and points within the shaded region are 1 standard
  deviation from the mean.}
\end{figure}

Fig.~\ref{defcomparison7}(a) shows scatter graphs of the optimal
values of the parameters for each vortex. The mean penetration depth
is $\Lambda_{ab}=\pdt{}$~nm and the standard deviation is
\SDpdt{}~nm. There are no obvious outliers in the scatter graph so
assuming that all the vortices have the same penetration depth gives
$\Lambda_{ab}=\pdt{} \pm \epdt{}$~nm.

To our surprise, the simplex algorithm gave thicknesses of
superconducting material which were considerably larger than the
calibrated values. The average calibrated thickness was \tc{}~nm but
the average fitted thickness was \ttt~nm. We were expecting the
fitted thicknesses to be shorter as there may have been damage introduced by
the focussed ion beam milling resulting in non-superconducting
material on the specimen surfaces. The problem seems to arise because
the specimen tilt and the specimen thickness are to some extent
complementary: if a vortex has a given projected length normal to the
electron beam, this could be the result either of a thick specimen
with a low tilt angle, $\alpha$, or a thin specimen with a high tilt
angle. The algorithm would then adjust the penetration depth to
account for this.

%Lab book 4/2/13 p143

To assess the extent to which a change in thickness changes the fitted
penetration depth, we fixed the thicknesses at their calibrated values
and re-ran the simplex algorithm. This resulted in
Fig.~\ref{defcomparison7}(b) which gave a mean penetration depth of
$\Lambda_{ab}=\pdc{}$~nm and a standard deviation of \SDpdc{}~nm so
that $\Lambda_{ab}=\pdc{} \pm \epdc{}$~nm. Fig.~\ref{defcomp7}(c)
shows that this change has almost no effect on the quality of the
fit. 

Finally we assess the effect of a dead-layer of non-superconducting
material on the surfaces of the sample. In addition to measuring the
total sample thickness via a combination of energy-filtered imaging
and holography as described in section~\ref{methods}, we also took
convergent-beam diffraction patterns from 5 regions of the sample under
two-beam conditions. These were used to find the thickness of the
crystalline component of the sample as described in
ref.~\onlinecite{Williams96}. The difference between the two
measurements gives the thickness of any amorphous layers present but
we found that to within the experimental error of $\pm 10$~nm, there
was no difference.

We repeated the simplex procedure with the superconducting thicknesses
fixed at 50~nm (72~nm parallel to the electron beam) below their
calibrated values which we consider to be the largest plausible
dead-layer thickness. Again, Fig.~\ref{defcomp7}(c) shows that the fit
is still good and this yielded an average penetration depth of
$\Lambda_{ab}=\pdd{}$~nm with a standard deviation of
\SDpdd~nm. Averaging over all the vortices gives $\Lambda_{ab}=\pdd{}
\pm \epdd{}$~nm. Thus we can say that at 10.8~K, $\Lambda_{ab}$ lies
between 100--130~nm and is more likely to be between 100--115~nm.

\section{Derivation of the Vortex Structure using the Transport of Intensity Equation}

In the previous section it was shown that simulated images based on
the London model of flux vortices in MgB$_2$ provided a good fit to
the experimental images. It would be advantageous, however, to derive
the phase and the B-field from the images directly so that no aspects
of the model are assumed. In principle, this can be done by
reconstructing the phase from the defocussed images using the
transport of intensity equation~\cite{Beleggia04}.

The transport of intensity equation is a re-expression of the
Schr\"{o}dinger equation in terms of the intensity, $I$, and phase,
$\phi$, of the electrons combined with the condition that there is a
constant flow of electrons. Using the same coordinate system as
section~\ref{comdefsim} where the electron beam approaches the
specimen in the $z$-direction, has wavelength $\lambda$ and the image
(located at $z=0$) has coordinates $x,y$, the transport of intensity
equation is:

\begin{equation}
\nabla_{xy}.\left(I\nabla_{xy}\phi\right)=-\left(2\pi \over \lambda\right){\partial I \over \partial z}
\end{equation}

%See lab book p80 18/12/12

When the in-focus image has a uniform intensity, $I_0$, (a good
approximation for the images acquired here) the equation can be
simplified to Poisson's equation:

\begin{equation}
\label{Poisson}
\nabla^2_{xy}\phi=-\left(2\pi \over \lambda I_0\right){\partial I \over \partial z}
\end{equation}

It can then be solved Fourier transforming in $x$ and $y$ but not $z$
to give the Fourier transform of the phase as:

\begin{equation}
\label{TIEFourier}
\widetilde{\phi}={1 \over 2\pi\lambda I_o}{1 \over k^2}\text{F.T.}\left[{\partial I \over \partial z}\right]
\end{equation}

where F.T. stands for the Fourier transform of the term in square
brackets. The rate of change of the intensity in the electron beam at
the exit-plane of the specimen $z=0$ may be estimated by taking two
images with equal and opposite defocus, $\pm\Delta f$, so that
${\partial I \over \partial z}\approx {I(\Delta f)- I(-\Delta f) \over
  2\Delta f}$. The phase in real-space can then be found by
calculating the inverse transform of Eqn.~\ref{TIEFourier}
numerically. The singularity at $k=0$ in Eqn.~\ref{TIEFourier} is
indicative of the fact that the zero of phase is arbitrary so a
constant can always be added without any physical consequences. Here
the pixel at $k=0$ was set to zero so the average phase in the image
was zero. It can also be seen from Eqn.~\ref{Poisson} that the
recovered phase is not just insensitive to an additive constant but
also to the addition of a phase ramp.

Ideally one would want to measure the B-field from the vortices. The
quantity closest to this which can be derived from electron
micrographs is the induction-thickness product, `$({\bf Bt})(x,y)$':
the component of the B-field normal to the electron beam integrated
along the length of the beam. It is related to the derivative of the
phase via:

\begin{equation}
({\bf Bt})(x,y)\equiv\int_{-\infty}^\infty \left(\begin{array}{c}B_x\\B_y\end{array}\right)\,{\rm d}z={h \over 2\pi e}\left(\begin{array}{c}-\partial\phi/\partial y\\\partial\phi/\partial x\end{array}\right)
\end{equation}

For a conventional magnetic material where the B-field is constant
through the thickness of the specimen and the stray field is
negligible, the induction-thickness product is simply the component of
the B-field normal to the electron beam multiplied by the thickness of
the specimen. In the case of flux vortices, the stray field is not
negligible and is included in the induction-thickness product. As
mentioned above, the transport of intensity equation is insensitive to
the addition of a phase ramp so the induction thickness product can
only be recovered to within an additive constant. As the simulations
were carried out with the same boundary conditions as the
reconstructions, both have the same constant offset.

\begin{figure}
\includegraphics[height=80mm]{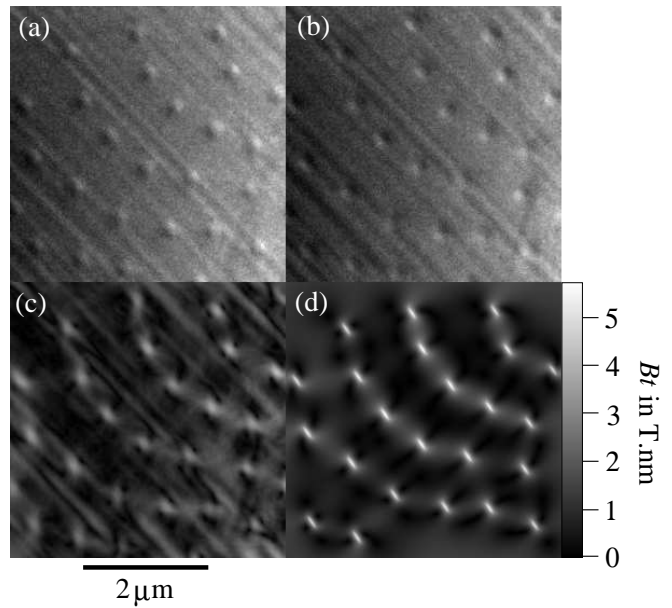}
\caption{\label{defocus} (a) Original image taken with $\Delta
  f=1.10\pm0.02$cm.(b) Original image taken with $\Delta
  f=-1.10\pm0.02$cm. (c) The magnitude of the induction-thickness
  product derived using the transport of intensity equation. (d)
  Simulated induction-thickness product for $\Lambda_{ab}=\pdc{}$~nm
  using the calibrated thicknesses.}
\end{figure}

Fig.~\ref{defocus} shows two images, equally disposed either side of
focus with $\Delta f=\pm 1.10$~cm. A longer defocus series was
obtained but the transport of intensity equation works best for images
closest to focus due to the approximation to the gradient of the
intensity with defocus explained earlier. From these images, the
modulus of the induction-thickness product shown in (c) was
derived. (d) shows a simulation of the induction-thickness product for
the same vortex array using $\Lambda_{ab}=\pdc{}$~nm and it can be
seen that it is much more strongly peaked at the centre of the
vortices. This could either indicate that the core of the vortices was
much wider than previously thought or that the transport of intensity
equation smooths the B-field profile.

\begin{figure}
\includegraphics{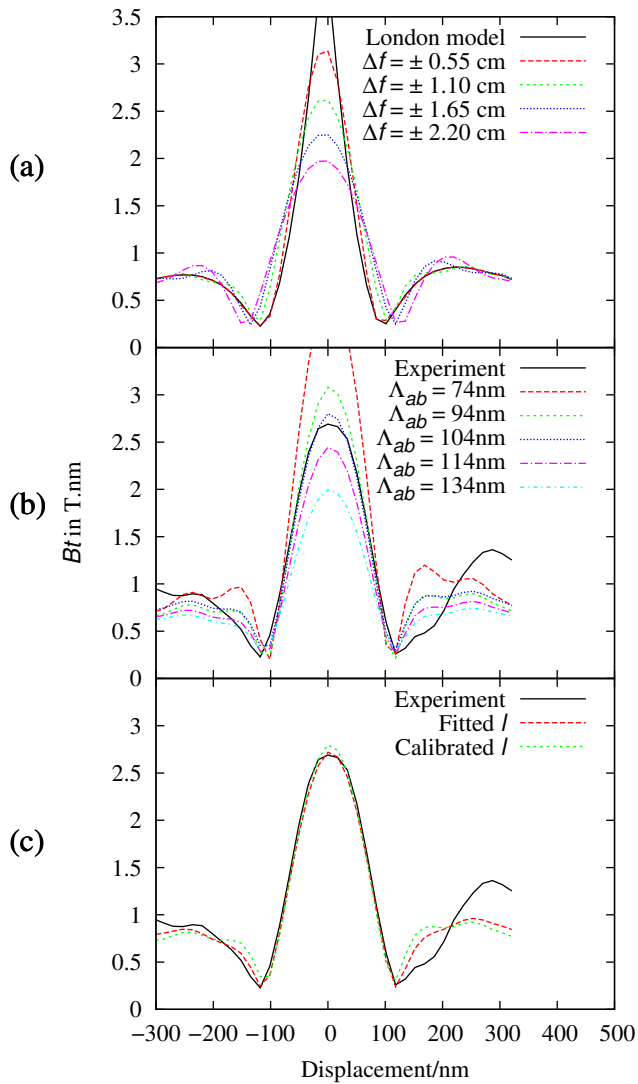}
\caption{\label{TIE} (a) The modulus of the induction-thickness
  product ($Bt$) for a simulated London vortex (solid line) and
  reconstructions based on pairs of simulated images at the defoci
  shown. $\Delta f=\pm 1.10$~cm was used in this experiment. (b)
  Comparison of the induction-thickness product ($Bt$) reconstructed
  from experimental data for a single vortex with simulations for
  different values of $\Lambda_{ab}$. (c) Induction-thickness profile
  derived from experimental data using the transport of intensity
  equation together with simulated profiles. In the first, the
  thickness parallel to the electron beam was fitted as part of the
  simplex routine yielding $l=530$~nm, $\alpha=52^\circ$,
  $\Lambda_{ab}=123$~nm. In the second simulation the thickness was
  fixed at the calibrated value of $l=373$~nm, yielding
  $\alpha=43^\circ$, $\Lambda_{ab}=104$~nm.  }
%Vortex 5 used.
\end{figure}

To ascertain which of these possibilities was correct, we used the
simulated phase shift to produce simulated defocussed images and then
applied the transport of intensity equation to these. Fig.~\ref{TIE}(a)
shows simulated profiles of the induction-thickness product across the
centre of a vortex and it can be seen that it is indeed the use of the
transport of intensity equation which broadens the B-field
profiles. This effect is caused by subtracting two images close to
focus to approximate the gradient ${\partial I \over \partial
  z}\approx {I(\Delta f)- I(-\Delta f) \over 2\Delta f}$ and not by a
lack of coherence in the electron beam as suggested in
ref.~\onlinecite{Beleggia04}.

Thus, the transport of intensity equation does not immediately yield
the B-field profile of a flux vortex. To compare experiment and
theory, it is first necessary to simulate defocussed images from the
theory and then generate a magnetic profile by applying the transport
of intensity equation to these. Fig.~\ref{TIE}(b) shows that when this
is done, a close match is obtained between the experimental profile
and simulations based on the London model and that if the penetration
depth were the only quantity subject to uncertainty, it would be
possible to determine a best fit to within $\pm 5$~nm.

As before, however, there is an ambiguity between the specimen
thickness and the specimen tilt angle and Fig.~\ref{TIE}(c)
shows that very similar profiles can be obtained for different sets of
parameters. The best values for the penetration depth were most likely
found by direct comparison between experimental and simulated images
in section~\ref{comdefsim} rather than the more convoluted process of
applying the transport of intensity equation to experimental images
and to simulated images and comparing the results. If this is done,
however, it yields $\Lambda_{ab}=\TIEpdt{} \pm \TIEepdt$~nm when the
thickness is treated as a variable parameter and
$\Lambda_{ab}=\TIEpdc{} \pm \TIEepdc$~nm when the thicknesses are
fixed at their calibrated values. As expected, these values are very
similar to those obtained by comparing the experimental images with
simulations in section~\ref{comdefsim}.

%%%%%%%%%%%%%%%%%%%%%%%%%%%%%%%%%%%%%%%%%%%%%%%%%%%%

\section{Summary and Conclusions}

Transmission electron microscopy gives a method to investigate the
magnetic structure of individual flux vortices in type-II
superconductors at low magnetic field and allows absolute measurements
of the penetration depth to be obtained from individual vortices
irrespective of whether they form a regular array. In
section~\ref{Fres_or_holo} we showed that out-of-focus imaging is a
more sensitive method for measuring the B-fields associated with flux
vortices compared with off-axis holography.

We investigated the magnetic structure of flux vortices in MgB$_2$ by
comparing simulations based on the London model of flux vortices with
out-of-focus images and found these gave a good fit indicating that
the London model gives a good description of flux vortices in
MgB$_2$. The main uncertainty in the fits was the fraction of the
sample thickness which was superconducting. Treating the thickness as
a variable parameter in the fits and averaging over \n{} vortices
yielded a penetration depth of $\Lambda_{ab}=\pdt{}\pm\epdt$~nm
although this gave an implausibly large specimen thickness: an average
of \ttt{}~nm as opposed to the calibrated average of \tc{}~nm. Fixing
the thicknesses at their calibrated values gave
$\Lambda_{ab}=\pdc{}\pm\epdc$~nm and reducing the thicknesses from
their calibrated thicknesses by 50~nm to mimic a dead layer of
non-superconducting material gave $\Lambda_{ab}=\pdd{}\pm\epdd$~nm
with only minimal changes to the goodness of fit. Thus at 10.8~K and
in a field of 27~Oe, $\Lambda_{ab}$ lies between 100--130~nm and is
more likely to be between 100--115~nm.

The most reliable meaurement of the penetration depth is likely to be
the small-angle neutron scattering experiments from single crystals by
Cubitt {\it et al.}~\cite{Cubitt03} which gave $\Lambda_{ab}=82 \pm
2$~nm at 2~K. According to the radio-frequency measurements of Manzano
{\it et al.}~\cite{Manzano02}, this will increase to $90 \pm 2$~nm at
10.8~K, the temperature at which the measurements were made
here. Although radio-frequency measurements are usually used to derive
changes in the penetration depth, Manzano {\it et
  al.}~\cite{Manzano02} were able to estimate absolute values of
$\Lambda_{ab}=110$--130~nm at 1.35~K which will increase to
$\Lambda_{ab}=118$--138~nm at 10.8~K using data from polycrystals and
values for the anisotropy. Our values lie between these two
measurements.

We also investigated the use of the transport of intensity equation
which in principle allows the magnetic fields in the specimen to be
derived without reference to a model. We found that although this gave
a useful overview of the B-fields from the vortices, the B-field
profiles are broadened due to the way the change of intensity with
defocus is approximated. This must be taken into account when
comparing the reconstructions with models so the technique does not
provide a model-free method to determine the magnetic structure of
vortices. It should be noted that if off-axis holography could be
used, the magnetic profile would also be broadened due to the size of
the Fourier mask used in the reconstruction process and this too would
need to be accounted for.

The coherence length $\xi_{ab}$ in MgB$_2$ is only
5.5~nm~[\onlinecite{Manzano02}] and we did not observe any rounding of
the B-field profile which could be ascribed to the coherence
length. For a material with a larger coherence length like niobium,
where the coherence length is 39~nm~[\onlinecite{Annett04}], observing
the effect on the B-field profile should be quite feasible.

Quantitative measurements of the magnetic structure of flux vortices
is only one aspect of the information which can be obtained using this
technique. Information on the pinning landscape of the specimen can be
derived from measurements of vortex motion~\cite{Sow98} and we are
currently preparing a paper on this subject.

%%%%%%%%%%%%%%%%%%%%%%%%%%%%%%%%%%%%%%%%%%%%%%%%%%%%

\begin{acknowledgments}
This work was funded by the Royal Society and the Engineering and
Physical Sciences Research Council, Grant No. EP/E027903/1. Work at
Eidgenössische Technische Hochschule Zürich was supported by the Swiss
National Science Foundation and the National Centre of Competence in
Research programme 'Materials with Novel Electronic Properties'.
\end{acknowledgments}

\section{Supplemental Information  -- The Effect of Noise on the Reconstructed Phase}

Here we investigate how the noise in an image is transferred to the
phase recovered from these images. The coordinate system is oriented
so that the electron beam travels along $z$ and the image lies in the
$xy$-plane.  For simplicity, we consider 1-dimensional images although
the results can be readily extended to 2-dimensions. The Fourier
transforms referred to are taken in $x$ and $y$ but not $z$.

\subsection{Noise in Phase Reconstructions from Holograms}

Here we show how noise present in an off-axis hologram is transferred
to the reconstructed phase. The standard expression~\cite{Volkl99} for
the intensity in a hologram from a pure phase object is given by

\begin{equation}
I(x)=1+V\cos\left(2\pi Qx+\phi(x)\right)+n(x)
\end{equation}

where $V$ is the visibility of the holographic fringes, $Q$ is the
carrier frequency, $\phi(x)$ is the true phase shift (as opposed to
that measured experimentally) and we have added noise, $n(x)$ which is
the noise-to-signal ratio for each pixel (note that the average
intensity per pixel has been normalised to 1). We now follow the usual
procedure to reconstruct the phase from the hologram.

The first step is to take the Fourier transform of the intensity of
the hologram which gives:

\begin{eqnarray}
\widetilde{I}(k)&=&\delta(k)+{V \over 2}\bigg(\text{F.T.}[e^{i\phi(x)}]*\delta(k-Q)\nonumber\\
&+&\text{F.T.}[e^{-i\phi(x)}]*\delta(k+Q)\bigg)+\widetilde{n}(k)
\end{eqnarray}

where F.T. denotes the Fourier transform of the term in square
brackets, * is a convolution and $\widetilde{n}(k)$ is the Fourier
transform of the noise. It can be seen that the Fourier transform is
composed of a central peak and two sidebands centred at $k=\pm Q$.

Next, the origin of the transform is shifted to the centre of the
sideband at $k=Q$ by convolving both sides of the above equation by
$\delta(k+Q)$. The sideband is then isolated by multiplying by a top
hat function $\widetilde{h}(k)$ which has the value 1 over the size of
the sideband and is zero elsewhere to give a new function
$\widetilde{J}(k)$:

\begin{equation}
\widetilde{J}(k)=\left({V \over 2}\text{F.T.}\left[e^{i\phi(x)}\right]+\widetilde{n}(k)*\delta(k+Q)\right)\widetilde{h}(k)
\end{equation}

The inverse transform gives: 

\begin{equation}
J(x)=\left({V \over 2}e^{i\phi(x)}+n(x)e^{2\pi iQx}\right)*h(x)
\end{equation}

\begin{figure}
\includegraphics[height=55mm]{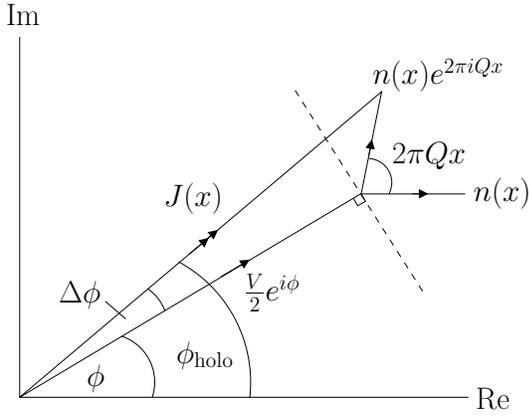}
\caption{\label{argand} Argand diagram showing the how noise in a
  hologram, $n(x)$, leads to noise $\Delta\phi$ in the recovered
  phase, $\phi$.}
\end{figure}

It can be seen that in the absence of noise, $J(x)$ is proportional to
the wavefunction of the electron beam as it exits the specimen,
$e^{i\phi}$. The inverse transform of the top-hat function, $h(x)$,
acts as a smoothing function and it controls the resolution of the
reconstructed phase. Referring to the Argand diagram in
Fig.~\ref{argand}, it can be seen that the phase measured by
holography at a particular position, $x$, is the sum of the true
phase, $\phi$ and an additional phase due to the noise $\Delta\phi$ so
that $\phi_{\text{holo}}=\phi+\Delta\phi$. The geometry of
Fig.~\ref{argand} shows that provided the noise-to-signal ratio is
much smaller than the fringe visibility, $n/V\ll 1$, the additional
phase shift is

\begin{equation}
\label{phase}
\Delta\phi(x)=2{n(x)\over V}\sin\left(2\pi Qx-\phi(x)\right)
\end{equation}

The noise in the phase recovered by holography is then the
root-mean-square value of $\Delta\phi(x)$ averaged over the
image. Denoting the average over the image by $\langle\rangle$, the
noise in the phase image is

\begin{equation}
\sigma_{\phi_{\text{holo}}}\equiv \sqrt{\langle(\Delta\phi)^2\rangle}=\sqrt{{2\over V^2}\langle n^2(x)\rangle}=\sqrt{2\over NV^2}
\end{equation}

The last expression applies if the noise is Shot noise and there are
$N$ counts per pixel. It is the same as the formula given by Lichte
{\it et al.}~\cite{Lichte87} although obtained by a different method.

Taking the $m$-th derivative (denoted by superscripts in brackets) of
Eqn.~\ref{phase} in the approximation that there are many pixels for
each holographic fringe, $X\ll 1/Q$ gives

\begin{equation}
\phi_{\text{holo}}^{(m)}(x)=\phi^{(m)}(x)+2{n^{(m)}(x)\over V}\sin\left(2\pi Qx-\phi(x)\right)
\end{equation}

The first derivative of the noise means the difference in the noise
between two neighbouring pixels divided by the size of the pixel, $X$:
$n'(x)={\left(n(x+X)-n(x)\right)/X}$. Subtracting the two noise terms
from adjacent pixels increases the average noise by $\sqrt{2}$ and so
the noise in the $m$-th derivative of the phase due to Shot noise in
the original hologram is:

\begin{equation}
\sigma_{\phi^{(m)}_{\text{holo}}}={1 \over X^m}\sqrt{2^{m+1} \over NV^2}
\end{equation}

%%%%%%%%%%%%%%%%%%%%%%%%%%%%%%%%%%%%%%%%%%%%%%%%%%%

\subsection{Noise in Phase Reconstructions from Fresnel Images}

Here we show how noise present in pairs of defocussed images is
transferred to the phase recovered using the transport of intensity
equation. The transport of intensity of equation is:

\begin{equation}
\nabla_{xy}.\left(I\nabla_{xy}\phi\right)=-\left(2\pi \over \lambda\right){\partial I \over \partial z}
\end{equation}

For a phase object with unit average intensity, this simplifies to

\begin{equation}
\nabla^2_{xy}\phi=-\left(2\pi \over \lambda\right){\partial I \over \partial z}
\end{equation}

The derivative is found from two images equally disposed either side
of focus so that

\begin{equation}
\nabla^2_{xy}\phi=-\left(2\pi \over \lambda\right){I(\Delta
  f)-I(-\Delta f) \over 2\Delta f}
\end{equation}

If $I$ denotes the ideal intensity of the image and $I_{\text{expt}}$
is the intensity measured experimentally subject to a noise-to-signal
ratio at each pixel of $n(x,y)$, we have

\begin{equation}
I_{\text{expt}}(\Delta f)=I(\Delta f)+n(\Delta f)
\end{equation}

and

\begin{equation}
I_{\text{expt}}(-\Delta f)=I(-\Delta f)+n(-\Delta f)
\end{equation}

Giving

\begin{equation}
\nabla^2_{xy}\phi_{\text{Fresnel}}=\nabla^2_{xy}\phi-{\pi\over\lambda\Delta f}\left(n(\Delta f)-n(-\Delta f)\right)
\end{equation}

As in the previous section, subtracting the two noise terms increases
the average noise by a factor of $\sqrt{2}$ so the noise in the $m$-th
derivative of the phase recovered by the transport of intensity
equation from images with an average count of $N$ electrons per pixel
and subject to Shot noise is

\begin{equation}
\sigma_{\phi^{(m)}_{\text{Fresnel}}}={\pi \over X^{m-2}\lambda\Delta f}\sqrt{2^{m-1} \over N}
\end{equation}

%%%%%%%%%%%%%%%%%%%%%%%%%%%%%%%%%%%%%%%%%%%%%%%%%%%%

\begin{acknowledgments}
This work was funded by the Royal Society and the EPSRC, Grant
No. EP/E027903/1. Work at ETH was supported by the SNSF and the NCCR
program MaNEP.
\end{acknowledgments}

\bibliography{MgB2}% Produces the bibliography via BibTeX.

\end{document}